\documentclass[letterpaper, 10 pt, conference]{ieeeconf}
\IEEEoverridecommandlockouts
  \usepackage{amsmath,amsfonts}\usepackage{amsmath,amsfonts}
\usepackage{amsmath,amsthm}
\usepackage{algorithmic}
\usepackage{algorithm}
\usepackage{array}
\usepackage{textcomp}
\usepackage{stfloats}
\usepackage{url}
\usepackage{verbatim}
\usepackage{graphicx}
\usepackage{graphbox}
\newenvironment{renum}
{\begin{enumerate}}
{\end{enumerate}}
\usepackage{cite}
\usepackage{xcolor}
\usepackage{hyperref}

\DeclareMathOperator*{\argmin}{argmin}

\DeclareMathOperator*{\subjectto}{subject \ to}

\begin{document}

\title{Safe and Stable Formation Control with Autonomous Multi-Agents Using Adaptive Control}

\author{Jose A. Solano-Castellanos$^{1}$, Peter A. Fisher$^{1}$, Anuradha Annaswamy$^{1}$ \thanks{$^{1}$ Department of Mechanical Engineering, Massachusetts Institute of Technology, Cambridge, MA 02139, USA. (email:{ \tt\footnotesize \{jsolanoc, pafisher, aanna\}@mit.edu}).
}
\thanks{This work was carried out using the support of the Boeing
Strategic University Initiative.}
}




\maketitle
\begin{abstract}
This manuscript considers the problem of ensuring stability and safety during formation control with distributed multi-agent systems in the presence of parametric uncertainty in the dynamics and limited communication. We propose an integrative approach that combines Adaptive Control, Control Barrier Functions (CBFs), and connected graphs. The main elements employed in the integrative approach are an adaptive control design that ensures stability, a CBF-based safety filter that generates safe commands based on a reference model dynamics, and a reference model that ensures formation control with multi-agent systems when no uncertainties are present. The overall control design is shown to lead to a closed-loop adaptive system that is stable, avoids unsafe regions, and converges to a desired formation of the multi-agents. Numerical examples are provided to support the theoretical derivations.
\end{abstract}



\section{Introduction}

Multi-agent systems (MAS) have received much attention because of their potential in completing tasks that a single agent could not complete efficiently on their own. Examples include exploration, surveillance, reconnaissance, rescue, and failure-tolerance, which occur in various problems related to motion planning and robotics. Typical problems in the context of MAS include consensus, formation control, coordination, and synchronization. This paper pertains to formation control.

Parametric uncertainties are inevitable in any problem where surveillance, reconnaissance, or rescue are involved. An important hallmark of autonomous MAS is the ability to self-tune the underlying controller in real-time so as to deliver the desired performance despite the presence of uncertainties. Yet another challenge that autonomous MAS face is the presence of constraints on states which may be due to obstacles, potential collisions, or other unsafe causes. It is important for the underlying controller to not only enable formation control but also ensure that any state constraints are met. Finally the presence of multi-agents may be distributed in nature. Therefore the underlying control solution must be realized using limited communication. We will address all of these features that may be present in MAS.

\begin{figure}[t!]
    \centering
    \includegraphics[width=0.95\columnwidth]{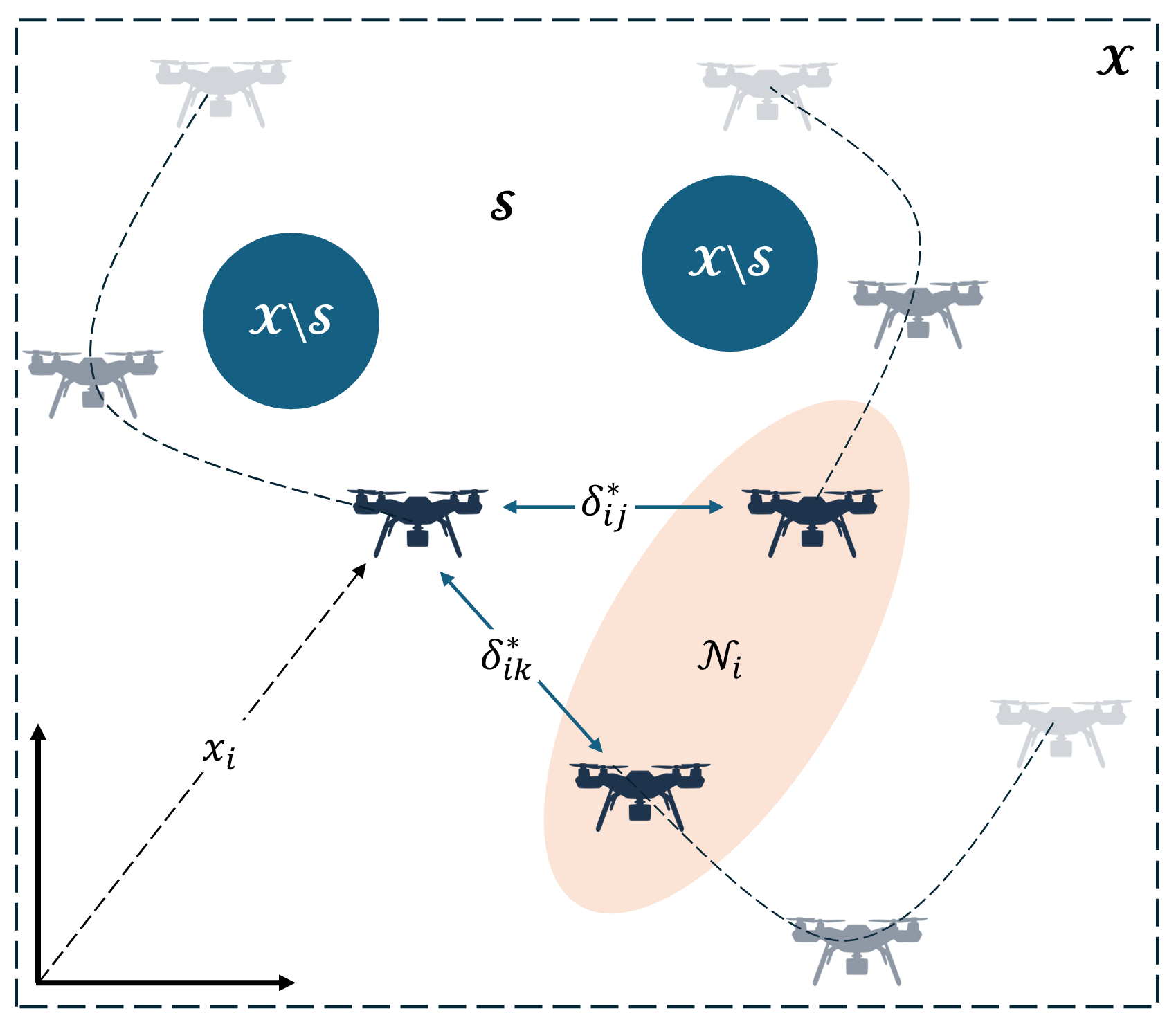}
    \caption{Graphical representation of the proposed method, where a multi-agent system with limited communication needs to converge to a desired formation while avoiding unsafe regions in the presence of parametric uncertainty in the model.}
    \label{fig:Splash}
\end{figure}

The concept of MAS formation control can be classified into the formation tracking and formation producing. In the former, the agents maintain a desired trajectory while the configuration itself moves through space, while in the latter, the objective is to converge to a static formation from some initial configuration of the agents, which is useful in tasks of surveillance and more generally sensor deployment. We will focus on a class of dynamic MAS with a goal of static formation using real-time control when subjected to parametric uncertainties, state-space constraints, and limited communication among the agents\footnote{In the sense that not all agents can communicate with every other agent.}.


Several approaches have been reported in the literature to achieve formation control, but with a subset of the above features. The approaches in \cite{Lavretsky2003AdaptiveNeurocontrol}, \cite{Dydek2013AdaptiveUAVs}, and \cite{Li2017MultipleAlgorithm} have addressed the formation control problem in the presence of parametric uncertainty using adaptive control. No constraints on the state or in the communication among agents are however considered. The solutions in \cite{Cai2015AdaptiveDynamics}, \cite{Xuan-Mung2019RobustApproach}, and \cite{Wang2014DistributedRobots} focus on formation control with limited communication and parametric uncertainties, but do not address obstacles or other state-space constraints. The approaches in \cite{Ge2022AAvoidance} and \cite{Shi2019AdaptiveSystems} address nonlinear dynamics as well, and propose a neural network based solution.
In contrast to the approaches in \cite{Taylor2020AdaptiveFunctions} and \cite{Lopez2021RobustSafety}, the safe and stable adaptive controller proposed here is less conservative, as the buffer introduced in the safety filter becomes arbitrarily small as adaptation proceeds. 

The control solution proposed in this paper advances the state of the art in formation control using MAS by considering a class of MAS with nonlinear dynamics, parametric uncertainties, state-space constraints, and limited communication. The controller includes adaptive control components, a safety filter based on Control Barrier Functions (CBFs), and a reference model that incorporates the topology with limited communication among the MAS. The overall closed-loop adaptive system is shown to be stable and safe with respect to specified state constraints,while achieving the desired formation. These rigorous analytical results are supported by simulation studies of a four-agent MAS, where our integrative safe and stable adaptive approach is required to successfully complete the formation task. The inclusion of all of these components including the analytical proofs represents the contribution of the paper.


Preliminaries are presented in Section \ref{Preliminaries} and the control problem in Section \ref{Control Problem}. The main contribution of this manuscript, the development of an adaptive controller with stability and safety properties, is presented in Section \ref{Adaptive Control}. Section \ref{Simulations} includes numerical simulations. Proofs of the main contributions of the manuscript (Theorems 2 and 3) are presented in the appendix.

\section{Preliminaries}\label{Preliminaries}

\subsection{Graph Theory and MAS Communication}

A graph $\mathcal{G}$ is a pair $\{\mathcal{V}, \mathcal{E}\}$ where $\mathcal{V} = \{1, \cdots, M\}$ is known as the vertex set which contains the vertices or nodes of the graph (i.e. the agents), and $\mathcal{E}$ is known as the edge set which contains a collection of pairs $(i,j)$ denoting the connectivity or communication between vertices $i$ and $j$ of the vertex set. The set of \textit{neighbors} of node $i$ is denoted as $\mathcal{N}_i \overset{\Delta}{=} \{ \ j \ | \ (i.j) \in \mathcal{E} \ \}$. 

A graph is called \textit{undirected} if node $i$ communicates with node $j$ and vice versa, then $(i,j) = (j,i)$. Otherwise, the graph is called \textit{directed} and node $i$ communicates with node $j$ but node $j$ does not necessarily communicate with node $i$, i.e. $(i,j) \neq (j,i)$. A graph is said to be \textit{connected} if it has a node to which there exists a (directed) path from every other node. For convenience of the following definitions, throughout this manuscript, only undirected graphs are considered. 

\textbf{Definition 1:} The degree matrix $D(\mathcal{G}) \in \mathbb{R}^{M \times M}$ of graph $\mathcal{G}$ is a diagonal matrix whose diagonal entries $d_{ii}$ are equal to the number of neighbors of agent $i$, i.e. $|\mathcal{N}_i|$\footnote{$|\cdot|$ denotes the cardinality of the set.}.

\textbf{Definition 2:} The adjacency matrix $A(\mathcal{G})\in \mathbb{R}^{M \times M}$ of graph $\mathcal{G}$ is a matrix whose diagonal entries are zero and the off diagonal entries $a_{ij}$ are given by
\begin{equation}\label{eq:aij}
    a_{ij} = \begin{cases} 
      1 & \text{if} \ (i,j) \in \mathcal{E} \\
      0 & \text{if} \ (i,j) \notin \mathcal{E}
   \end{cases}
\end{equation}

\textbf{Definition 3:} The Laplacian matrix $L(\mathcal{G})\in \mathbf{S}^M_{+}$ of graph $\mathcal{G}$ is defined as
\begin{equation}\label{eq:laplacian}
    L(\mathcal{G}) \overset{\Delta}{=} D(\mathcal{G}) - A(\mathcal{G})
\end{equation}

Notice that the sum of the rows and columns of the Laplacian matrix add up to zero, therefore $\alpha \mathbf{1}\in\mathbb{R}^M$, for all $\alpha \in \mathbb{R}$, lies on the right and left null space of $L$. Furthermore, if the graph is connected, the Laplacian has $M-1$ non-zero eigenvalues, i.e. $\lambda_M \geq \cdots \geq \lambda_2 > \lambda_1 = 0$ \cite{Merris1994LaplacianSurvey}.

\textit{Notation:} $\mathbf{S}_{+}^n$ and  $\mathbf{S}_{++}^n$ denote the symmetric positive semidefinite and definite cones of dimension $n$, respectively. $\lambda_{max}(C)$ and $\lambda_{min}(C)$ denote, respectively, the largest and smallest eigenvalues of a symmetric matrix $C$.

\subsection{Control Barrier Functions (CBFs)}

Consider a nonlinear control system that is affine in control
\begin{equation}\label{affine}
    \dot{x} = f(x) + g(x)u
\end{equation}
where $x \in \mathcal{X} \subset \mathbb{R}^n$, $f: \mathcal{X} \rightarrow \mathbb{R}^n$ and $g: \mathcal{X} \rightarrow \mathbb{R}^n$ are locally  Lipschitz, and $u \in \mathcal{U} \subseteq \mathbb{R}^m$. Safety can be defined in terms of a continuously differentiable function $h : \mathcal{X} \rightarrow \mathbb{R}$ and a set $\mathcal{S} \subset \mathcal{X}$, such that
\begin{subequations}\label{eq:safety}
    \begin{gather}
        \mathcal{S} \overset{\Delta}{=} \{ \ x \in \mathcal{X} \ | \ h(x) \geq 0 \ \}\\
        \partial\mathcal{S} \overset{\Delta}{=} \{ \ x \in \mathcal{X} \ | \ h(x) = 0 \ \}\\
        \text{int} (\mathcal{S}) \overset{\Delta}{=} \{ \ x \in \mathcal{X} \ | \ h(x) > 0 \ \}
    \end{gather}
\end{subequations}

The notion of a CBF \cite{Prajna2004SafetyCertificates, Ames2014ControlControl} can be formulated such that its existence allows the system in (\ref{affine}) to be rendered safe with respect to $\mathcal{S}$ in the sense that the set is made weakly positively invariant for some input $u$. The following definition formalizes this notion.

\textbf{Definition 4 \cite{Ames2014ControlControl}}: Let $\mathcal{S \subset \mathcal{X}}$ be the zero-superlevel set of $h : \mathcal{X} \rightarrow \mathbb{R} $. The function
$h$ is a \textit{Zeroing Control Barrier Function} (ZCBF) for $\mathcal{S}$, if there exists an extended class kappa function, $\alpha (h(x)) \in \mathcal{K_\infty}$, such that for the system (\ref{affine}) it can be obtained that:
\begin{equation}\label{eq:safety_u}
    \sup_{u\in\mathcal{U}} \left[ L_fh(x) + L_gh(x)u\right] \geq -\alpha (h((x)))
\end{equation}
for all $x \in \mathcal{S}$, where $L_\chi h$ is the Lie derivative of $h$ with respect to $\chi$.

We now introduce the following definition that pertains to the safety of MAS.

\textbf{Definition 5 \cite{Zhang2023NeuralControl}:} A continuously differentiable function $h: [\mathcal{X}]^{|\mathcal{N}_i|+1} \rightarrow \mathbb{R}$ is referred as a Graph Control Barrier Function (GCBF) if there exists $\alpha(h(x_i,\mathbf{x}_{\mathcal{N}_i})) \in \mathcal{K}_\infty$ and a control law $k_{i}: [\mathcal{X}]^{|\mathcal{N}_i|+1} \rightarrow \mathcal{U} \subseteq \mathbb{R}^m$, such that
\begin{equation}\label{eq:GCBF}
    \sum_{j \in \{i,\mathcal{N}_i\}}\left[L_fh(x_j) + L_gh(x_j)u_j \right] \geq -\alpha(h(x_i,\mathbf{x}_{\mathcal{N}_i})) 
\end{equation}
for all $i \in \mathcal{V}$ and $u_j = k_j(x_i(t),\mathbf{x}_{\mathcal{N}_i}(t))$ where $\mathbf{x}_{\mathcal{N}_i}$ is 
the joint state of the agents in neighbor set $\mathcal{N}_i$.

For ease of exposition, the time dependency will not be made explicit going forward unless a new variable is introduced or it is relevant to the presented argument.


\section{The Control Problem}\label{Control Problem}

The problem we consider is this paper is the static formation control  of distributed MAS consisting of $M$ agents indexed by $i \in \mathcal{V} = \{1, \ \cdots, M\} $, in the presence of parametric uncertainties and obstacles, with the goal of ensuring stability and safety. The dynamics of agent $i$ are given by
\begin{equation}\label{eq:Dynamics}
    \dot{x}_i = Ax_i + B\left[\Lambda u_i + \Xi \omega\right]
\end{equation}
where $x_i(t) \in \mathbb{R}^n$ is the state of the agent and $u_i(x(t),t) \in \mathbb{R}^m$ is the control input of the agent. $A \in \mathbb{R}^{n \times n}$ is unknown, $\Lambda \in \mathbb{R}^{m \times m}$ is an unknown diagonal matrix with known sign, and $B \in \mathbb{R}^{n \times m}$  is known and it has full column rank. The term $\Xi \omega$ corresponds to nonlinearities present in the system where $\omega(x_i(t),t) \in \mathbb{R}^p$ is known, but $\Xi \in \mathbb{R}^{m \times p}$ is unknown. Several nonlinearities in practical problems take the form $\Xi \omega$: see \cite{Slotine1991AppliedControl} for examples in robotic systems and \cite{Tong2011Observer-BasedSystems} for an example in chemical reactors. Similar nonlinearities also occur in Radial Basis Function Neural Networks (see for example, \cite{Ge2022AAvoidance} and \cite{Shi2019AdaptiveSystems}).
For ease of exposition, we assume that $A, \ \Lambda$ \text{and} $\Xi$ are independent of $i$; extensions to the case when they depend on $i$ are straightforward. We introduce the following assumptions regarding the unknown parameters and nonlinearities in the dynamics:

\textbf{Assumption 1:} The matrix $\text{sign}(\Lambda)\Lambda$ is diagonal and positive definite, $\text{sign}(\Lambda)\Lambda \in \mathbf{S}_{++}^m$.

\textbf{Assumption 2:} The nonlinearity $\omega(x_i(t),t)$ is a bounded signal, i.e. $||\omega(x_i(t),t)||_{\infty} \leq \bar{\omega}$ for all $t \geq t_0$.

\textbf{Assumption 3:} A constant matrix $\theta^{\star}_1 \in \mathbb{R}^{m \times n}$ exist such that
\begin{equation}\label{eq:theta1}
    A + B\Lambda\theta^{\star}_1 = A_m
\end{equation}
where $A_m \in \mathbb{R}^{n \times n}$ is a known Hurwitz matrix (Section \ref{Reference Model}).

In addition to the parametric uncertainties, we assume that the problem has safety considerations in the form of state constraints. The state constraints  are captured by the safe set $S_i$ for each agent $i \in \mathcal{V}$ which corresponds to the obstacle-free region of the state space. Associated with this safe set is a GCBF $h(x_i)$, which implies the existence of a $u_i$ in (\ref{eq:Dynamics}) that guarantees (\ref{eq:GCBF}). We introduce the following assumption regarding the safe set $S_i$. 

\textbf{Assumption 4:} The safe set $S_i \subset \mathcal{X}$ is bounded and connected.


It is important to note that inter-agent collisions safety is not considered in this paper. Nonetheless, this could be incorporated into our solution by using the decentralized safety barrier constraints introduced by \cite{Wang2017SafetySystems}.

The overall problem statement is therefore the following: given that the initial condition $x_i(t_0) \in \mathcal{S}_i$ for all $i \in \mathcal{V}$ and a desired position in the static formation $x_i^{\star} \in \mathcal{S}_i$ for all $i \in \mathcal{V}$, the problem is to find a control policy of the form
\begin{equation}\label{eq:ControlProblem}
    u_i = k_i(x_i,x_{\mathcal{N}_i})
\end{equation}
where $k_i : [\mathbb{R}^n]^{|\mathcal{N}_i|+1} \rightarrow \mathbb{R}^m$, that guarantees that $x_i(t) \in \mathcal{S}_i$ for all $ t \geq t_0$, and  $x_i(t) \rightarrow x_i^{\star}$ in the presence of unknown parameters. 


\section{A Safe and Stable Adaptive Controller}\label{Adaptive Control}

The controller that we propose includes an adaptive component together with a safety filter. In order to guide the adaptive control design, a reference model is suitably chosen. In Section \ref{Stable}, we design such a reference model, the corresponding adaptive controller, and show that the adaptive controller can enable the MAS to reach a desired formation. In Section \ref{SafeandStableFormationControl}, we  integrate a safety filter into the adaptive control design and show that the desired formation can be reached even while ensuring the safety constraints. The corresponding results are stated in Theorem 2, Corollary 1, Corollary 2, and Theorem 3.

\subsection{Stable Formation Control}\label{Stable}

\subsubsection{Reference Model}\label{Reference Model} The starting point for the adaptive solution to the MAS formation control is the choice of a reference model which specifies the desired trajectory that the MAS should follow. For this purpose, we propose a reference model dynamics similar to that in \cite{Olfati-Saber2004ConsensusTime-delays, Li2010ConsensusViewpoint}:
\begin{subequations} \label{eq:Reference_i}
    \begin{gather}
        \dot{z}_i = A_mz_i + B \bar{u}_i\\
        \bar{u}_i = \Theta\left[\sum_{j \in \mathcal{N}_i}(z_j-z_i) - \Delta_{i}\right]
    \end{gather}
\end{subequations}
where $z_i(t) \in \mathbb{R}^n$ is the state of the reference model, $A_m \in \mathbb{R}^{n\times n}$ is a Hurwitz matrix, $\Delta_{i}(t) \in \mathbb{R}^n$ is the reference input vector, and $\Theta \in \mathbb{R}^{m \times n}$ is a constant control gain. The choice of $\Delta_i$ is dictated by the static formation that is of interest. The gain $\Theta$ ensures that the closed-loop reference system remains stable, for a given graph $\mathcal{G}$.

The following theorem specifies the conditions under which the reference model (\ref{eq:Reference_i}a), (\ref{eq:Reference_i}b) leads to the desired formation \cite{ Li2010ConsensusViewpoint}. 

\textbf{Theorem 1 \cite{ Li2010ConsensusViewpoint}}: If the graph that captures the communication among agents ($\mathcal{G}$) is connected, i.e. $\lambda_2 > 0$, then a choice of $\Theta$
\begin{subequations}\label{eq:Theta}
\begin{gather}
    \Theta = c B^TR^{-1}\\
    c \geq 1/\lambda_2  
\end{gather}
\end{subequations}
where $R \in \mathbf{S}^n_{++}$ is the solution of
\begin{equation}\label{eq:Matching}
    A_m^TR + RA_m - 2B B^T + Q_0 = 0
\end{equation}
for a given $Q_0 \in \mathbf{S}^n_{++}$, and a choice of the reference input
\begin{equation}\label{eq:ref_input}
    \Delta_{i}(t) = \sum_{j \in \mathcal{N}_i}\delta_{ij}^{\star}
\end{equation}
for all $t \geq t_0$, where $\delta_{ij}^{\star} \overset{\Delta}{=} x_i^{\star} - x_j^{\star}$, ensures that $z_i(t)$ converges to the desired formation $x_i^{\star}$, for all $i \in \mathcal{V}$. 

\noindent \textbf{Remark 1:} The result of Theorem 1 holds if the communication graph ($\mathcal{G}$) varies with time as long as it remains connected.

\noindent \textbf{Remark 2:} The main point to note here is that if the underlying MAS dynamics is in the form of (\ref{eq:Reference_i}a), where $A_m$ and $B$ are known, a controller of the form of (\ref{eq:Reference_i}b) guarantees that the states of (\ref{eq:Reference_i}a) will converge to the desired formation. However, we note that the MAS dynamics (as shown in (\ref{eq:Dynamics})) has parametric uncertainties, as $A$, $\Lambda$, and $\Xi$ are unknown. We therefore are unable to ensure that the control input (\ref{eq:Reference_i}b) suffices. We therefore propose an adaptive controller for this problem below, and forms the first contribution of this paper.

\subsubsection{Adaptive Controller}\label{OnlyAdapatation}
Since the MAS dynamics in (\ref{eq:Dynamics}) has nonlinearities, with unknown $A$ (and therefore $\theta_1^\star$), $\Lambda$ and $\Xi$, we introduce an adaptive control input
\begin{equation}\label{eq:control_i}
    u_i = \hat{\theta}^{(i)}_1x_i + \hat{\theta}^{(i)}_2\left[\sum_{j \in \mathcal{N}_i}(x_j-x_i) - \Delta_{i}\right] + \hat{\theta}^{(i)}_3 \omega(x_i,t)
\end{equation}
where $\hat{\theta}^{(i)}_k(t)$
for $k = \{1, 2, 3\}$ are time-varying parameters that are adjusted. The adjustable parameters $\hat{\theta}_k^{(i)}$ are introduced for different purposes: $\hat{\theta}^{(i)}_1$ is utilized for stabilizing the linear dynamics, $\hat{\theta}^{(i)}_2$ is used to enable convergence to the desired formation, while $\hat{\theta}^{(i)}_3$ is used to compensate for the nonlinearities. We define the following unknown parameters:
\begin{equation}\label{eq:theta2}
    \theta_2^{\star} \overset{\Delta}{=}  -\Lambda^{-1}\Theta
\end{equation}
\begin{equation}\label{eq:theta4}
    \theta_3^{\star} \overset{\Delta}{=} -\Lambda^{-1}\Xi
\end{equation}
where $\theta^{\star}_2 \in \mathbb{R}^{m \times n}$ and $\theta^{\star}_3 \in \mathbb{R}^{m \times p}$. Choosing $\hat{\theta}_k^{(i)}=\theta_k^*$, for all $i = \mathcal{V}$ and $k= \{ 1, 2, 3\}$, guarantees that the closed-loop system specified by the plant in (\ref{eq:Dynamics}) and the controller in (\ref{eq:control_i}) matches the reference model in (\ref{eq:Reference_i}a)-(\ref{eq:Reference_i}b). We also note that (\ref{eq:control_i}) follows the communication constraint specified in (\ref{eq:ControlProblem}), i.e. $\bar{u}_i = k_i(z_i,z_{\mathcal{N}_i})$, where $z_{\mathcal{N}_i}$ is the joint state of the reference model agents in neighbor set $\mathcal{N}_i$. We propose the adaptive laws:
\begin{subequations}\label{eq:adaptation}
    \begin{gather}
        \dot{\hat{\theta}}^{(i)}_1 = -\Gamma_1 \Omega_i x_i^T\\
        \dot{\hat{\theta}}^{(i)}_2 = -\Gamma_2 \Omega_i \left[\sum_{j \in \mathcal{N}_i}(x_j-x_i) - \Delta_{i}\right]^T\\
        \dot{\hat{\theta}}^{(i)}_3 = -\Gamma_3 \Omega_i \omega^T
    \end{gather}
\end{subequations}
where $\Omega_i \overset{\Delta}{=} \text{sign}(\Lambda)B^TP(x_i - z_i)$, $\Gamma_k \in \mathbf{S}^m_{++}$ for all $k = \{1, 2, 3\}$, and a $Q \in \mathbf{S}^n_{++}$ 
is chosen so that
\begin{subequations}\label{eq:P}
    \begin{gather}
        A_m^TP + PA_m + Q = 0\\
        R^{-1}BB^TP + PBB^TR^{-1} \succeq 0
    \end{gather}
\end{subequations}
where $R$ is given by (\ref{eq:Matching}a).

\begin{figure*}[!t]
\centering
\includegraphics[width=0.675\textwidth]{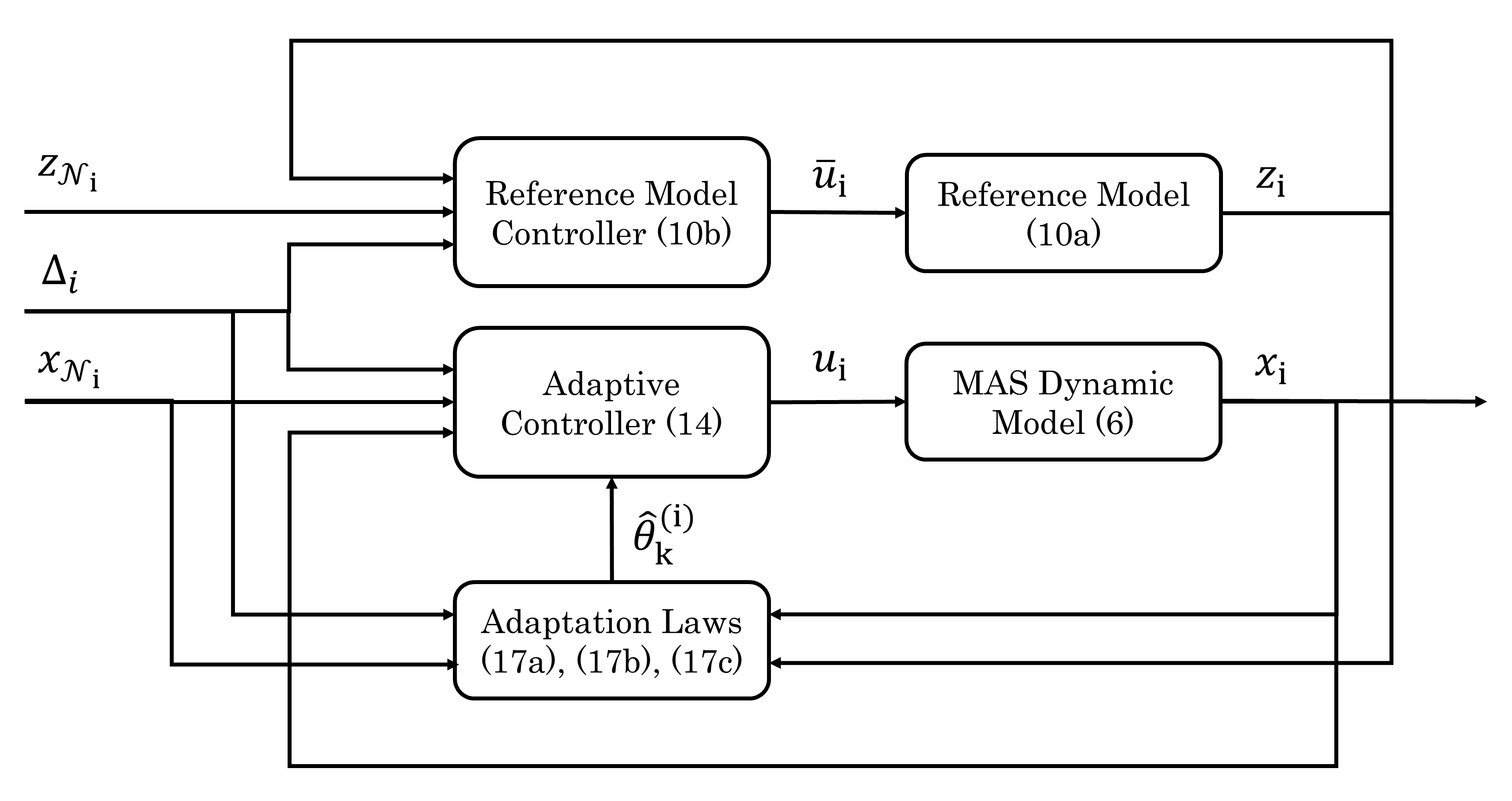}
\caption{Schematic of the proposed adaptive controller for stable formation control of MAS.}
\label{fig:Adaptive_Block_Diagram}
\end{figure*}

The overall adaptive controller schematic can be visualized in Fig. \ref{fig:Adaptive_Block_Diagram}. We now state the first main result of the paper, whose proof can be found in the appendix.

\textbf{Theorem 2}: The closed-loop adaptive system defined by the agent dynamics in (\ref{eq:Dynamics}), the control input (\ref{eq:control_i}) and the adaptation laws in (\ref{eq:adaptation}) has globally bounded solutions for any initial conditions $x_i(t_0)$ and $\hat{\theta}^{(i)}_k(t_0)$, for all $i \in \mathcal{V}$ and $k = \{1, 2,3\}$. Furthermore $x_i(t)$ converges to the desired final position $x_i^{\star}$, for all $i \in \mathcal{V}$, as $t \rightarrow \infty$. 

\begin{proof}
    See proof in Appendix \ref{appendix A}.
\end{proof}

\noindent \textbf{Remark 3:} The solution for $P$ in (\ref{eq:P}a)-(\ref{eq:P}b) can be obtained using LMI approaches, with $Q$ and $Q_0$ as decision variables.

We now introduce an important property of this adaptive controller which helps us integrate the CBF-filter in to the control design in order to ensure safety constraints on the state. We define the following variables
\begin{equation}
    \mathbf{e}(t) \overset{\Delta}{=} \mathbf{x}(t) - \mathbf{z}(t)
\end{equation}
\begin{equation}
    \mathbf{z}(t) \overset{\Delta}{=} [z_1^T \ \cdots \ z_M^T]^T
\end{equation}
\begin{equation}
    \mathbf{x}(t) \overset{\Delta}{=} [x_1^T \cdots \ x_M^T]^T
\end{equation}
\begin{equation}
    \tilde{\mathbf{\theta}} \overset{\Delta}{=} \left[\tilde{\theta}_1^{(1)} \ \tilde{\theta}_2^{(1)} \ \tilde{\theta}_3^{(1)} \ \cdots \ \tilde{\theta}_1^{(M)} \ \tilde{\theta}_2^{(M)} \ \tilde{\theta}_3^{(M)}\right]
\end{equation} 
\begin{equation}
    \tilde{\theta}^{(i)}_k(t) \overset{\Delta}{=} \hat{\theta}^{(i)}_k(t) - \theta^{\star}_k, \qquad \forall i \in \mathcal{V}, \ k = \{1, 2, 3\}
\end{equation}
\begin{equation}
    u_i^{\star} \overset{\Delta}{=} 
    \theta^{\star}_1x_i + \theta^{\star}_2\left[\sum_{j \in \mathcal{N}_i}(x_j-x_i) - \Delta_{i}\right] + \theta^{\star}_3 \omega
\end{equation}
\begin{equation}\label{eq:u_tilde_i}
\begin{split}
    \tilde{u}_i & \overset{\Delta}{=} u_i - u_i^{\star} \\
    & = \tilde{\theta}^{(i)}_1x_i + \tilde{\theta}^{(i)}_2\left[\sum_{j \in \mathcal{N}_i}(x_j-x_i) - \Delta_{i}\right] + \tilde{\theta}^{(i)}_3 \omega
\end{split}
\end{equation}
\begin{equation}
    \tilde{\mathbf{u}}(t) \overset{\Delta}{=} [\tilde{u}_1^T \ \cdots \ \tilde{u}_M^T]^T
\end{equation}
where $\mathbf{z}(t)$ is the joint state of the reference model, $\mathbf{x}(t)$ is the joint state of the MAS, $\tilde{\theta}^{(i)}_k(t)$ are the parameter errors, and $\tilde{u}_i$ are the input errors measured relative to the input with perfect knowledge of the parameters, $u_i^{\star}$. 

We then define bounds $C_\theta$, $C_\Lambda$, and $C_e$ given by $||\tilde{\mathbf{\theta}}(t_0)||_F \leq C_{\theta}$, $||\Lambda||_F \leq C_{\Lambda}$ and $||\mathbf{e}(t_0)|| \leq C_e$. With these bounds,  we  introduce the following corollaries to Theorem 2:

\textbf{Corollary 1:} For any bounded $\delta_{ij}^{\star}$, the following properties of $\mathbf{e}(t)$ hold:
\begin{renum}
    \item For all $t \geq t_0$, $||\mathbf{e}(t)|| \leq E$ where 
    \begin{equation}\label{eq:ErrorBound}
        E = \sqrt{\frac{\lambda_{max}(P)C_e^2 + \gamma^{-1}C_{\Lambda}C_{\theta}^2}{\lambda_{min}(P)}}
    \end{equation}
    and $\gamma \overset{\Delta}{=} \max(\lambda_{max}(\Gamma_1), \lambda_{max}(\Gamma_2), \lambda_{max}(\Gamma_3))$.
    \item For any $\varepsilon_e > 0$, there exists a finite time $T_e$ such that, $||\mathbf{e}(t)|| \leq \varepsilon_e$, for all $t \geq t_0 + T_e$
\end{renum}
\begin{proof}
    See proof in Appendix \ref{appendix B}. 
\end{proof}
\textbf{Corollary 2:} For any bounded $\delta_{ij}^{\star}$, the following properties of $\tilde{u}_i(t)$ hold: 
\begin{renum}
    \item $\tilde{u}_i \in \mathcal{L}_{\infty}$.
    \item For any $\varepsilon_u > 0$, there exists a finite time $T_u$ such that, $||\tilde{\mathbf{u}}(t)|| \leq \varepsilon_u$, for all $t \geq t_0 + T_u$
\end{renum}
\begin{proof}
    See proof in Appendix \ref{appendix C}. 
\end{proof}

\noindent \textbf{Remark 4:} Corollaries 1(i) and 2(i) state that $\mathbf e(t)$ and $\tilde{\mathbf{u}}(t)$ are always bounded, while Corollaries 1(ii) and 2(ii) state that $\mathbf e(t)$ and $\tilde{\mathbf{u}}(t)$ become arbitrarily small after a finite time $T=\max\{T_e,T_u\}$. The corresponding bounds $E$, $\varepsilon_e$, and $\varepsilon_u$ will be directly leveraged in designing the safety-inducing CBF filter introduced in the next section.

\noindent \textbf{Remark 5:} We note that the bound $E$ depends on $C_{\theta}$ and $C_{\Lambda}$, which in turn can be estimated from known upper bounds on the unknown parameters $\Lambda$ and $\theta^{(i)\star}_k$, for all $i \in \mathcal{V}$ and $k \in \{1,2,3\}$. This in turn implies that a known bound $E$ can be determined. Such a bound will be utilized in the GCBF introduced in the next section.

\noindent \textbf{Remark 6:} We note that this complete section is devoted to addressing stability of the closed-loop adaptive system and realizing a desired formation. We did not address the safety properties of the MAS. In particular, we did not evaluate the behavior of the closed-loop states $x_i$ in relation to the safe set $S_i$. This is the focus of the next section.

\subsection{Safe and Stable Formation Control}\label{SafeandStableFormationControl}
In order to ensure the safety of the closed-loop dynamics, we first consider safety of the reference dynamics in (\ref{eq:Reference_i}a), (\ref{eq:Reference_i}b). That is, we want to ensure that $z_i(t) \in \mathcal{S}_i$ for all $t \geq t_0$ if $z_i(t_0) \in \mathcal{S}_i$. A choice of a QP-GCBF filter inspired by \cite{Autenrieb2023SafeSystemsb}, ensures this safety, and is specified as the solution of the following optimization problem:
\begin{equation}\label{eq:ZCBF_i}
\begin{aligned}
    \argmin_{\Delta_i} & \quad ||\Delta_i - \Delta^{\star}_i||_2^2 \\
    \subjectto & \quad \frac{\partial h}{\partial z_i}^T\dot{z}_i \geq -\alpha_0 h(z_i) + \upsilon(t)
\end{aligned}
\end{equation}
where
\begin{equation}\label{eq:ref_input_star}
    \Delta_{i}^{\star} \overset{\Delta}{=} \sum_{j \in \mathcal{N}_i}\delta_{ij}^{\star}
\end{equation}
$\alpha_0 > 0$ is any finite constant, $\upsilon(t) > 0$ for all $t\geq t_0$ is a safety buffer, and $h$ is a GCBF that is a function of the \textit{i-th} agent state and the parameters of the obstacle. We denote the solution of (\ref{eq:ZCBF_i}),(\ref{eq:ref_input_star}) as  $\Delta_i^{\psi}$.

\noindent \textbf{Remark 7:} It is important to note that the QP-GCBF filter above can be solved in a decentralized manner. This is because the lumped quantity $\Delta_i$ is used as the decision variable instead of the relative distances $\delta_{ij}$, and because the GCBF $h$ depends solely on the \textit{i-th} agent's state and the obstacle parameters. Together, they allow a decentralized solution to the QP-GCBF filter.

\noindent \textbf{Remark 8:} Notice that the addition of the safety buffer, $\upsilon(t)$, to the GCBF defines an implicit safe set given by
\begin{equation}
\tilde{\mathcal{S}}_i(\upsilon(t)) \overset{\Delta}{=} \left\{ \ x_i \in \mathcal{X} \ \left|  \ h(x_i) +  \frac{\upsilon(t)}{\alpha_0} \geq 0 \right. \right\}
\end{equation}
Also, notice that as $\upsilon(t) \rightarrow 0$, $\tilde{\mathcal{S}}_i(\upsilon(t)) \rightarrow \mathcal{S}_i$.

We now proceed to address the safety of the actual closed-loop dynamics of (\ref{eq:Dynamics}) with the adaptive controller given by (\ref{eq:control_i}),(\ref{eq:adaptation}a)-(\ref{eq:P}b). For this purpose, we note three points:
\begin{enumerate}
    \item The addition of a safety filter as in (\ref{eq:ZCBF_i}),(\ref{eq:ref_input_star}) produces a new reference ($\Delta_i^{\psi}$) rather than the one in (\ref{eq:ref_input}).
    \item Corollary 1 implies that the adaptive system state $x_i(t)$ approaches the reference state $z_i(t)$ and therefore $h(x_i(t))$ approaches $h(z_i(t))$ as $t \rightarrow \infty$, and
    \item We note that
    \begin{equation}
        |h(x_i(t))-h(z_i(t))| \leq \kappa_1 E
    \end{equation}
    \begin{equation}
        \left|\left|\left.\frac{\partial h(w)}{\partial w}\right|_{w=x_i(t)} -\left.\frac{\partial h(w)}{\partial w}\right|_{w=z_i(t)}\right|\right|_{\infty} \leq \kappa_2 E
    \end{equation}
    for all $t \geq t_0$ and all $i \in \mathcal{V}$, where $\kappa_1$ and $\kappa_2$ are the (local) Lipschitz constants of $h$ and $\frac{\partial h(w)}{\partial w}$.
\end{enumerate}
A safety filter can now be designed for the closed-loop adaptive system. In order to account for the difference between the system state and the reference state for all time $t\geq t_0$, we define the safety buffer in (\ref{eq:ZCBF_i}) as follows:
\begin{equation}\label{eq:upsilon_t}
    \upsilon(t) =\begin{cases} 
      \bar{\eta}(E,\bar{U}_i(t),t), & t_0 \leq t < t_0 + \hat{T} \\
      \bar{\eta}(\varepsilon_e, \varepsilon_u,t) & t \geq t_0 + \hat{T} 
   \end{cases}
\end{equation}
where
\begin{equation}\label{eq:eta_bar}
\begin{split}
    \bar{\eta}(\xi_e,\xi_u,t) = &\alpha_0\kappa_1\xi_e + \kappa_2\xi_e\left|\left|A_m z_i(t) + B \bar{u}_i(t)\right|\right|_{\infty} + \\
    &\left|\frac{\partial h(x_i(t))}{\partial x_i}^T\right|\left(\xi_e \bar{A} \mathbf{1}_n + C_\Lambda\xi_uB\mathbf{1}_m\right)
\end{split}
\end{equation}
such that $\bar{A} \in \mathbb{R}^{n \times n}$ is an element-wise upper bound of 
$A$\footnote{Meaning $(A)_{ij} \leq (\bar{A})_{ij}$, for all $i, j = \{1, \cdots, n\}$.}, $\bar{U}_i(t)$ is an upper bound for $\tilde{u}_i$, which can be computed as
\begin{equation}\label{eq:U_bar}
    \bar{U}_i(t) = C_{\theta}\left|\left|x_i + \sum_{j \in \mathcal{N}_i}(x_j-x_i) - \Delta_{i}^{\psi}(t_-) + \omega(x_i) \right|\right|_{\infty}
\end{equation}
where $\Delta_{i}^{\psi}(t_-)$ is the previous solution of the QP-GCBF and $\hat{T} \geq T = \max\{T_e,T_u\}$ is an estimate of the the time it takes $||\mathbf{e}(t)|| \leq \varepsilon_e$ and $||\tilde{\mathbf{u}}(t)|| \leq \varepsilon_u$ for given $\varepsilon_e$ and $\varepsilon_u$, for $t \geq t_0 + T$. 

The complete integrative safe and stable adaptive controller is given by (\ref{eq:adaptation}a), (\ref{eq:adaptation}c), (\ref{eq:P}a), (\ref{eq:P}b), a control input for the MAS
\begin{equation}\label{eq:QP_solution}
    u_i = \hat{\theta}^{(i)}_1x_i + \hat{\theta}^{(i)}_2\left[\sum_{j \in \mathcal{N}_i}(x_j-x_i) - \Delta_{i}^{\psi}\right] + \hat{\theta}^{(i)}_3 \omega
\end{equation}
a control input for the reference model,
\begin{equation}\label{eq:mod_reference_model_u}
   \bar{u}_i = \Theta\left[\sum_{j \in \mathcal{N}_i}(z_j-z_i) - \Delta_{i}^{\psi}\right] 
\end{equation}
and the adaptive law
\begin{equation}\label{eq:theta_dot_2_mod}
    \dot{\hat{\theta}}^{(i)}_2 = -\Gamma_2 \Omega_i \left[\sum_{j \in \mathcal{N}_i}(x_j-x_i) - \Delta_{i}^{\psi}\right]^T
\end{equation}

\begin{figure*}[!t]
\centering
\includegraphics[width=0.725\textwidth]{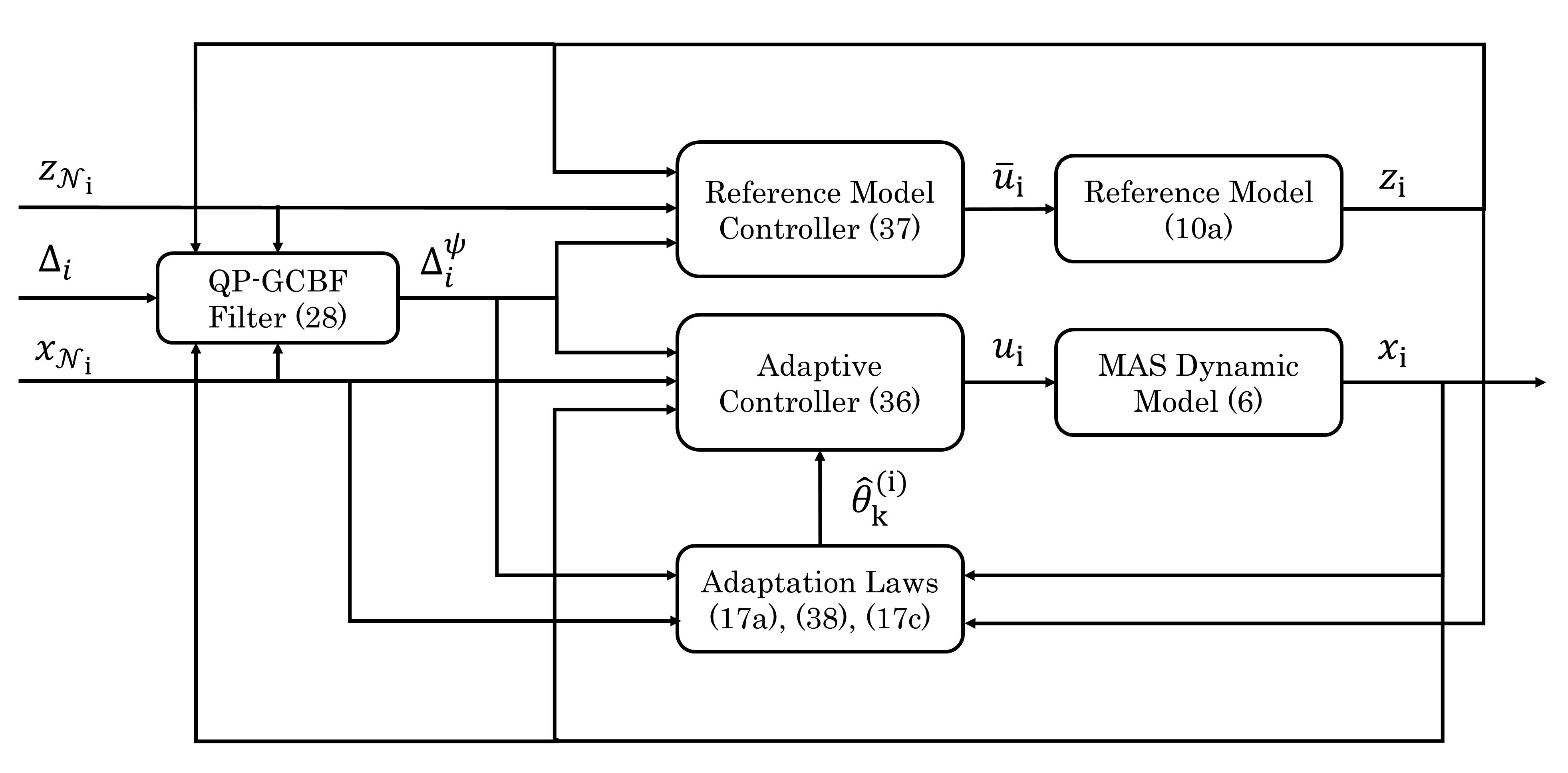}
\caption{Schematic of the proposed adaptive controller for safe and stable formation control of MAS.}
\label{fig:SafeAdaptive_Block_Diagram}
\end{figure*}

The schematic of the overall adaptive system with safety constraints can be visualized in Fig. \ref{fig:SafeAdaptive_Block_Diagram}. We now state the second main result of this paper.

\textbf{Theorem 3}: The closed-loop adaptive system defined by the plant in (\ref{eq:Dynamics}), the reference model (\ref{eq:Reference_i}a) and (\ref{eq:mod_reference_model_u}), the solution of the modified QP-GCBF in (\ref{eq:ZCBF_i}),(\ref{eq:ref_input_star}), and (\ref{eq:upsilon_t})-(\ref{eq:U_bar}), and the adaptive controller given by (\ref{eq:QP_solution}), (\ref{eq:adaptation}a), (\ref{eq:theta_dot_2_mod}), (\ref{eq:adaptation}c), (\ref{eq:P}a) and (\ref{eq:P}b) guarantees that:
\begin{renum}
    \item $x_i(t) \in \mathcal{S}_i$ for all $t \geq t_0$, if $x_i(t_0)\in \mathcal{S}_i$ for all $i \in \mathcal{V}$.
    \item Furthermore, for a given \(\varepsilon > 0\) such that the implicit safe set \(\tilde{\mathcal{S}}_i(\varepsilon)\) is connected, if \(\varepsilon_e\), \(\varepsilon_u\), and \(\hat{T}\) are chosen so that \(\upsilon(t) \leq \varepsilon\) for \(t \geq t_0 + \hat{T}\), then \(x_i(t) \to x_i^{\star}\) as \(t \to \infty\), provided that \(x_i^{\star} \in \tilde{\mathcal{S}}_i(\varepsilon)\) for all \(i \in \mathcal{V}\).
\end{renum}

\begin{proof}
    See proof in Appendix \ref{appendix D}. 
\end{proof}

\noindent \textbf{Remark 9:}  It is clear that knowing $\hat{T}$ is a requirement of our proposed approach. This in turn implies that the designer has a knowledge of how quickly transients decay during adaptation. It should be noted that it is guaranteed that such a $\hat{T}$ exists as the adaptive system is proved to be uniformly asymptotically stable.

\noindent \textbf{Remark 10:} Notice that for a given formation $x_i^{\star} \in \mathcal{S}_i$, there exists a sufficiently small $\varepsilon$ such that $x_i^{\star} \in \tilde{\mathcal{S}}_i(\varepsilon)$. This comes at the cost of a larger time window \([t_0, t_0+\hat{T}]\), during which the system operates more conservatively.

\noindent \textbf{Remark 11:} We make a few comments comparing the result in Theorem 3 with those in \cite{Taylor2020AdaptiveFunctions} and \cite{Lopez2021RobustSafety}. In the first paper, the authors define a piecewise continuous safety filter which tends to pull the plant away from the boundary of the safe set when it gets too close, creating an effective buffer window around the boundary. Although the size of this buffer window is independent on the size of the parameter error, the piecewise nature of the safety filter introduces undesirable switching behavior. In the second paper, the authors remedy this issue by introducing a persistent buffer window around the boundary of the safe set which is dependent on an upper bound on the magnitude of the parameter error. In fact, the term $\frac{1}{2}\tilde{\vartheta}^\top\Gamma^{-1}\tilde{\vartheta}$ in \cite{Lopez2021RobustSafety} is analogous to the quantity $\upsilon(t)$ in our paper, and both introduce significant conservatism. The advantage of our approach is that, due to the convergence of the plant to the reference model, for any arbitrarily small $\varepsilon_e, \varepsilon_u > 0$, there exists a finite time $T=\max\{T_e,T_u\}$ beyond which we can reduce our conservatism from $E$ to $\varepsilon_e$ and $\bar{U}_i(t)$ to $\varepsilon_u$ \textit{without any requirements on excitation or parameter learning}. 

The benefit of using the reference model dynamics rather than the plant dynamics in the GCBF conditions is our ability to accomplish simultaneous control and safety with only one adaptive law. If the GCBF conditions are determined in terms of the plant with an adaptive law to account for the uncertainties, then when control objectives (e.g. tracking) are introduced, a separate adaptive law will be needed to accomplish that goal. For instance, \cite{Taylor2020AdaptiveFunctions} uses both an adaptive Control Lyapunov Function (aCLF) and an adaptive Control Barrier Function (aCBF), with two separate adaptive laws. In contrast, our GCBF conditions are independent of the uncertainties, meaning that one adaptive law accomplishes both objectives.

\section{Simulation}\label{Simulations}

The proposed controller is applied to a two-dimensional obstacle avoidance problem. The dynamics of each agent are given by
\begin{equation}
    \left[\begin{matrix} \dot{x}_i\\
    \dot{y}_i
    \end{matrix}\right] = \left[ \begin{matrix} 0.25 & 0\\
    0 & 0.25
    \end{matrix}\right]\left[\begin{matrix} x_i\\
    y_i
    \end{matrix}\right] + \left[ \begin{matrix} 1.5 & 0\\
    0 & 1.5
    \end{matrix}\right]\left[\Lambda u_i + \Xi \omega\right]
\end{equation}
such that $x_{i}$ and $y_{i}$ are respectively the horizontal and vertical positions of the agents, actuation has been compromised ($\Lambda = 0.7I$) and the nonlinearities are characterized by 
\begin{equation}
    \Xi = \left[ \begin{matrix} 0.25 & -0.1\\
    -0.1 & 0.25
    \end{matrix}\right]
\end{equation}
\begin{equation}
        \omega(x_i,y_i,t) = \left[ \text{sign}(x_i)\cos(5t) \ \text{sign}(y_i)\sin(5t)\right]^T
\end{equation}

\begin{figure}
    \centering    \includegraphics[width=0.5\columnwidth]{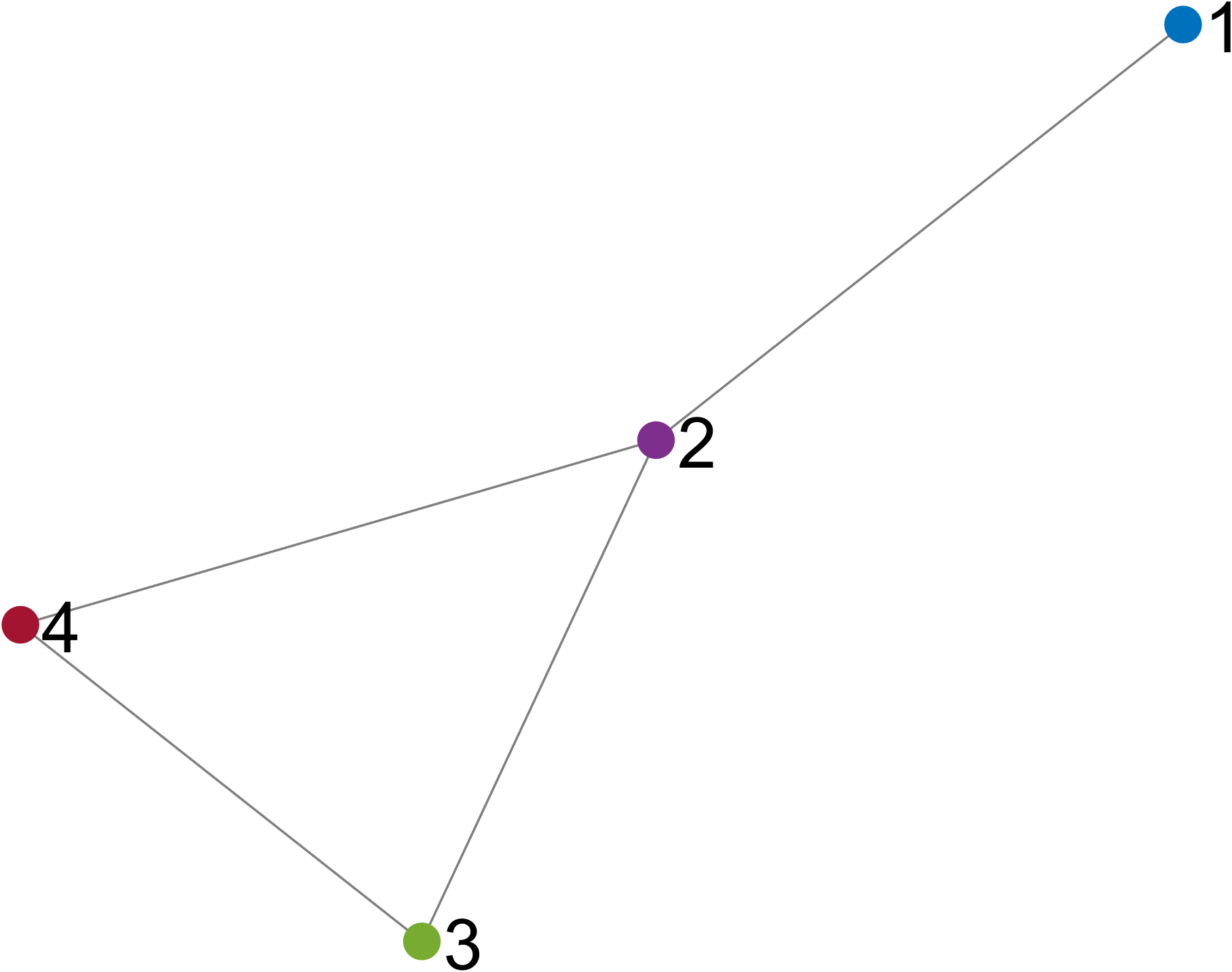}
    \caption{MAS communication graph for the obstacle avoidance example.}
    \label{fig:Graph}
\end{figure}

The reference model is defined by
\begin{equation}
    A_m = \left[ \begin{matrix} -2 & 0\\
    0 & -2
    \end{matrix}\right]
\end{equation}
and the communication graph is shown in Fig. \ref{fig:Graph}. To guarantee the safety of each agent, the following CBF constraint is chosen for each obstacle
\begin{equation}
    h_l(x_i,y_i) = (x_{i}-\bar{x}_{l})^2 + (y_{i}-\bar{y}_{l})^2 - r_l^2 \geq 0
\end{equation}
where $(\bar{x}_{l}, \bar{y}_{l})$ and $r_l$ are the position and radius of the \textit{l-th} circular obstacle.

The parameters $\hat{\theta}_k^{(i)}$, for all $i \in \mathcal{V}$ and $k = \{1,2,3\}$, are initialized assuming the agents dynamics are given by
\begin{equation}
    \left[\begin{matrix} \dot{x}_i\\
    \dot{y}_i
    \end{matrix}\right] = \hat{A}\left[\begin{matrix} x_i\\
    y_i
    \end{matrix}\right] + \left[ \begin{matrix} 1.5 & 0\\
    0 & 1.5
    \end{matrix}\right]\left[\hat{\Lambda} u_i + \hat{\Xi} \omega\right]
\end{equation}
where $\hat{A}=0.16I$, $\hat{\Lambda} = I$, and
\begin{equation}
    \hat{\Xi} = \left[ \begin{matrix} 0.35 & -0.15\\
    -0.15 & 0.35
    \end{matrix}\right]
\end{equation}

\begin{figure}
    \centering    \includegraphics[width=0.9\columnwidth]{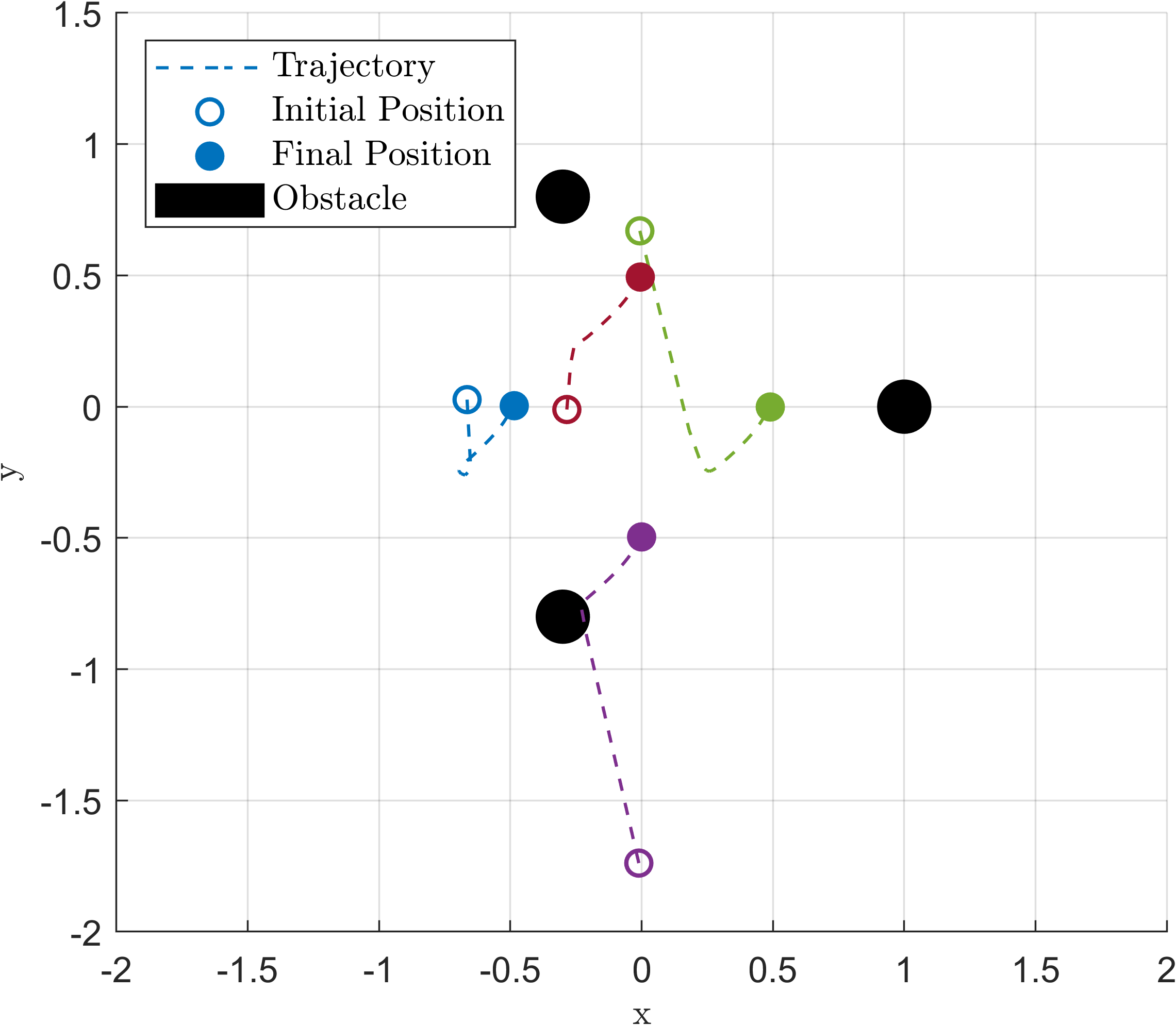}
    \caption{Trajectory of the MAS in the presence of obstacles, using only the adaptive controller.}
    \label{fig:AdaptationOnly}
\end{figure}

\begin{figure}
    \centering    \includegraphics[width=0.9\columnwidth]{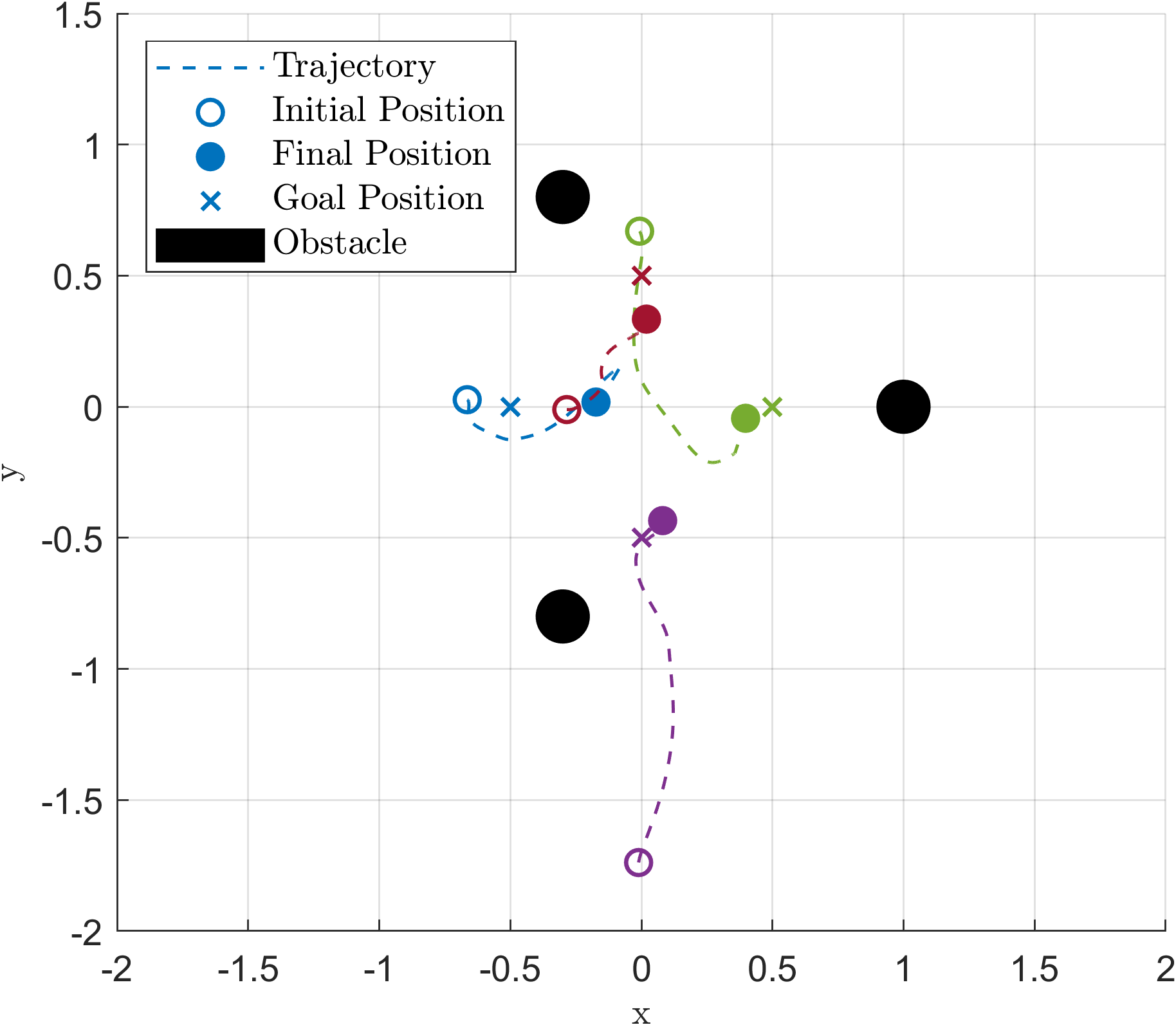}
    \caption{Trajectory of the MAS in the presence of obstacles, using only the safety filter.}
    \label{fig:CBF}
\end{figure}

\begin{figure}
    \centering
    \begin{minipage}{0.48\textwidth}
        \centering
        \includegraphics[width=\linewidth]{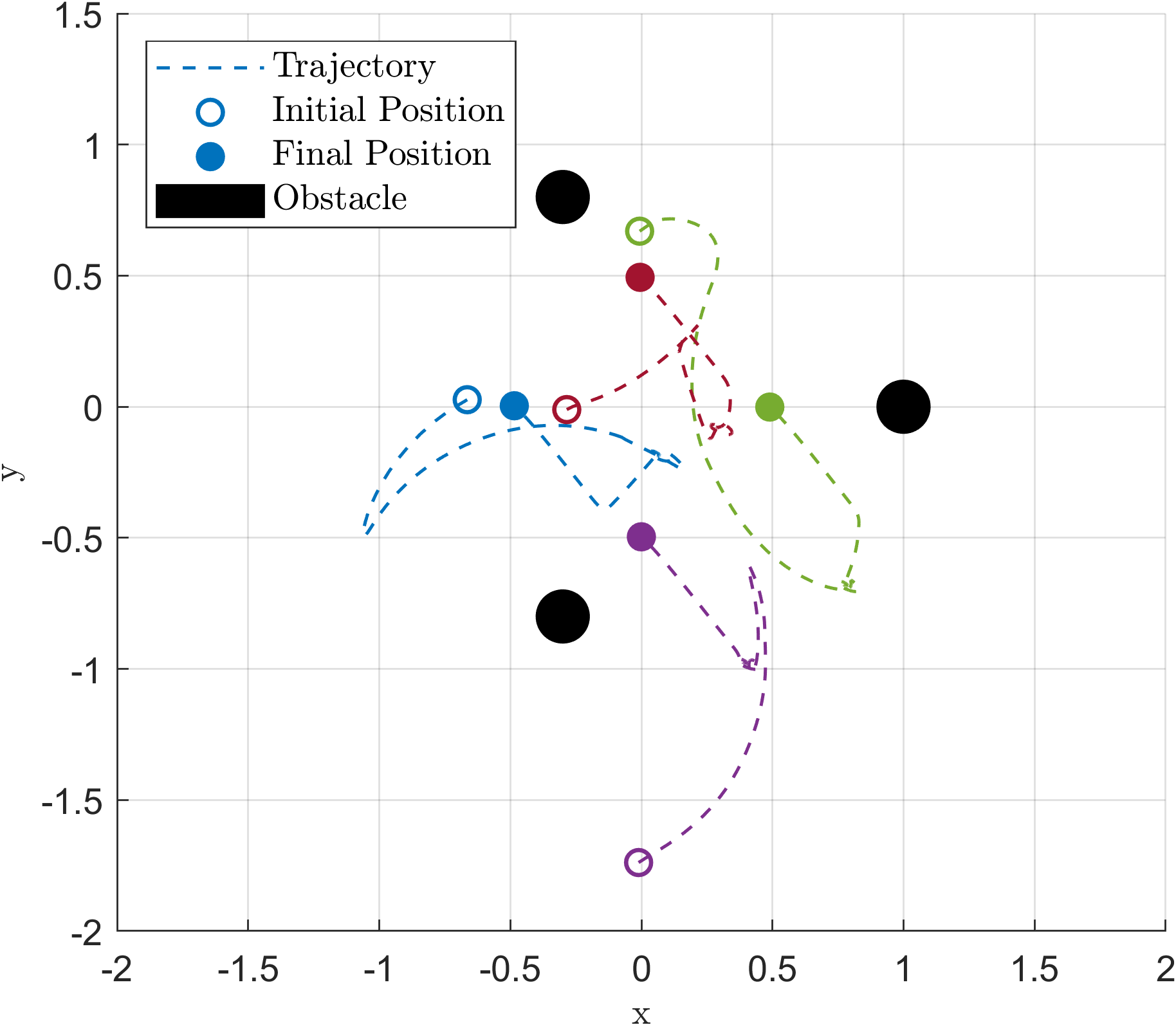}
        \\ (a)
    \end{minipage}
    \hfill
    \begin{minipage}{0.48\textwidth}
        \centering
        \includegraphics[width=\linewidth]{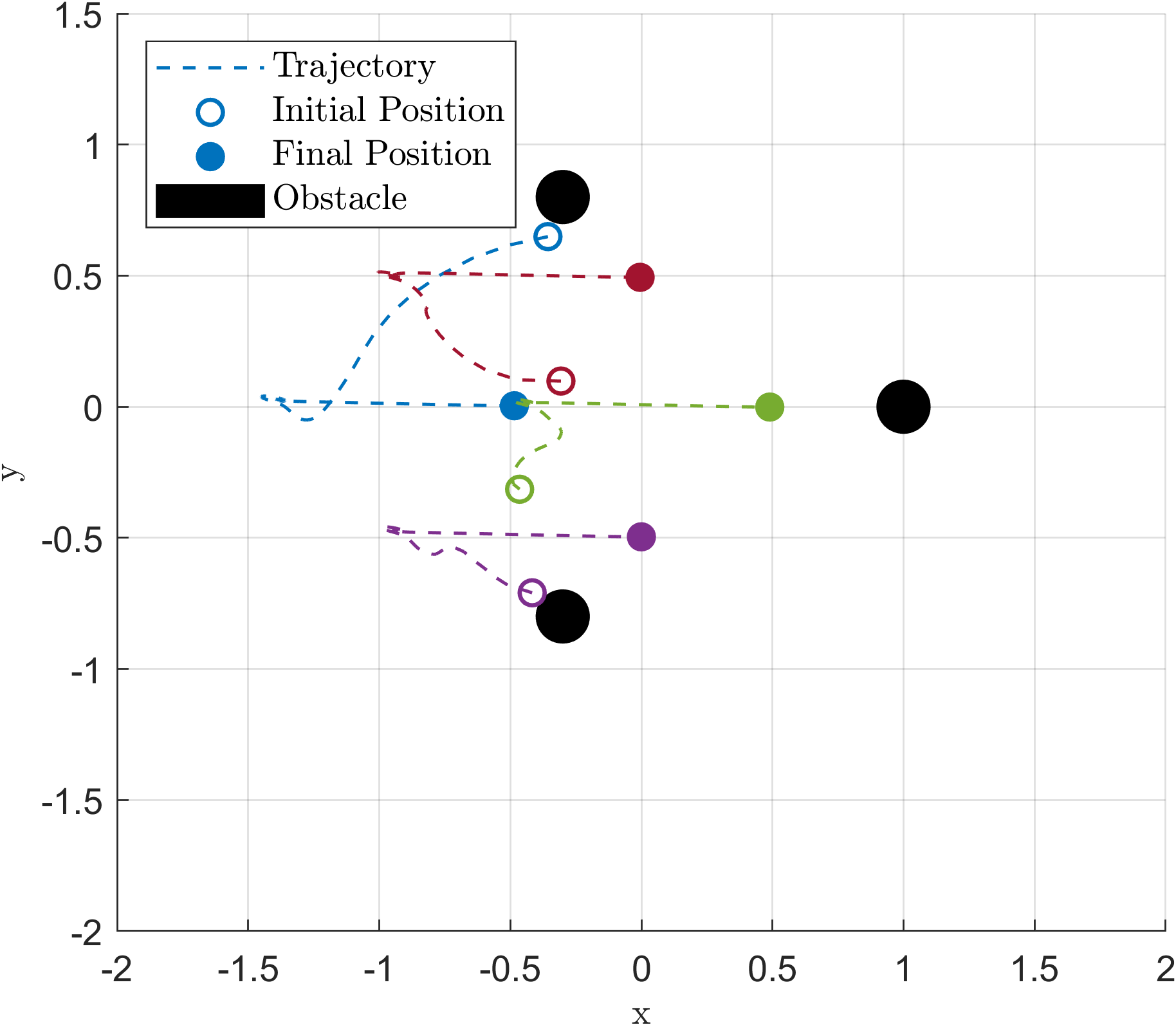}
        \\ (b)
    \end{minipage}
    \caption{Trajectory of the MAS in the presence of obstacles, using the proposed adaptive controller with control barrier functions.}
    \label{fig:ACBF}
\end{figure}

Fig. \ref{fig:AdaptationOnly} shows the trajectories of the agents using only the adaptive control  described in Section \ref{OnlyAdapatation} without any safety filter.  Fig. \ref{fig:CBF} shows the trajectories of the agents with the QP-GCBF filter in (\ref{eq:ZCBF_i}) but without adaptation. Finally, Fig. \ref{fig:ACBF} shows our proposed integrative adaptive controller with the QP-GCBF and the graph-based reference model as in (\ref{eq:Reference_i}). For the simulations that use the QP-GCBF,
\begin{equation}\label{eq:upsilon_sim}
    \upsilon(t) =\begin{cases} 
      0.5, & t_0 \leq t < t_0 + \hat{T} \\
      0.05 & t \geq t_0 + \hat{T} 
   \end{cases}
\end{equation}
$\hat{T}=5 \ s$, and $\alpha_0 = 15$ were chosen. The superior performance of our proposed controller compared with adaptation but no safety as well as safety but with no adaptation is clear from these figures. When the adaptive controller is employed but without any GCBF, agent 2 collides with an obstacle along the trajectory (Fig. \ref{fig:AdaptationOnly}). Only using the QP-GCBF but without adaptation allows the MAS to remain safe, but the agents do not reach the desired formation (Fig. \ref{fig:CBF}).  In contrast, with our proposed approach, agents reach the desired formation without collision (Fig. \ref{fig:ACBF}). Notice that even when the agents start very close to the obstacles, our approach is able to safely direct the MAS to the desired formation (Fig. \ref{fig:ACBF}b). We refer the reader to \href{http://aaclab.mit.edu/autonomous-flight-systems.php}{aaclab.mit.edu} for animations of the results. It was observed that a larger $\upsilon(t)$, that can be determined using known upper bounds on the parametric uncertainties (\ref{eq:eta_bar}),(\ref{eq:U_bar}), led to a more conservative performance with the trajectories staying far from the obstacle during the transient phase of the adaptation.

It's crucial to emphasize that the system successfully accomplishes the task without collisions, despite operating under unstable dynamics and partially known nonlinearities. Furthermore, it manages to do so even when the actuation is compromised in an unknown manner.

\section{Conclusions}\label{Conclusions}
In this paper, we consider the problem of static formation control with distributed MAS in the presence of parametric uncertainties and limited communication. The class of problems considered is nonlinear systems that are feedback linearizable, with states accessible for measurement. The goal is to ensure that the MAS stay inside a safe set with the overall closed-loop system remaining stable while meeting the formation goals. Our approach is a combination of adaptive control and CBFs, with the former providing a means for accommodating to parametric uncertainties and the latter providing a safety filter that ensures that the states stay within a safe region and remain forward-invariant. The innovations are the design of a QP-GCBF filter to account for parametric uncertainties and the design of a graph-based reference model which serves as desired dynamics that ensures a safe formation. Theoretical results are provided that guarantee global boundedness  and safety against obstacles in the overall state space, and convergence to the desired formation. Numerical results show the effectiveness of the proposed method.

Future work will focus on extensions to higher relative degree in the underlying MAS models, and in relaxing the requirement of knowing $\hat{T}$.

\bibliographystyle{ieeetr}

\appendix

\subsection{Theorem 2}\label{appendix A}
\begin{proof} 
We rewrite the reference model in (\ref{eq:Reference_i}a)-(\ref{eq:Reference_i}b) as 
\begin{subequations}\label{eq:Reference}
    \begin{gather}
    \mathbf{\dot{z}} = \left[ (I \otimes A_m) - (L\otimes B\Theta) \right]\mathbf{z} - (I \otimes  B\Theta)\Delta \\
    \Delta = \left[\Delta_1^T \ \cdots \  \Delta_M^T\right]^T
    \end{gather}
\end{subequations}
where $\otimes$ is the Kronecker product. The error dynamics can be written as:
\begin{equation}\label{eq:error_dynamics}
    \mathbf{\dot{e}} = \left[ (I \otimes A_m) - (L\otimes B\Theta) \right]\mathbf{e} + \Psi_1\mathbf{x} + \Psi_2 + \Psi_3
\end{equation}
where
\begin{equation}
    \Psi_1(t) = \text{diag}\left(B\Lambda\tilde{\theta}^{(1)}_1, \cdots, B\Lambda\tilde{\theta}^{(M)}_1 \right)
\end{equation}
\begin{equation}
    \Psi_2(t) = 
    \left[ \begin{matrix} B\Lambda\tilde{\theta}^{(1)}_2\left[\sum_{j \in \mathcal{N}_1}(x_j-x_1) - \Delta_{1}\right]\\
    \vdots\\
    B\Lambda\tilde{\theta}^{(M)}_2\left[\sum_{j \in \mathcal{N}_M}(x_j-x_M) - \Delta_{M}\right]
    \end{matrix} \right]
\end{equation}
\begin{equation}
    \Psi_3(t) = 
    \left[ \begin{matrix} B\Lambda\tilde{\theta}^{(1)}_3\omega(x_1, t)\\
    \vdots\\
    B\Lambda\tilde{\theta}^{(M)}_3\omega(x_M, t)
    \end{matrix} \right]
\end{equation}

We consider the following Lyapunov function candidate:
\begin{equation}\label{eq:Lyapunov}
    V = \mathbf{e}^T\mathbb{P}\mathbf{e} + \sum_{k=1}^3\sum_{i=1}^M \text{Tr}\left( \tilde{\theta}^{(i)T}_k\Gamma_k|\Lambda|\tilde{\theta}^{(i)}_k \right)
\end{equation}
where $\mathbb{P} = I \otimes P \in \mathbf{S}^{Mn}_{++}$, since $P \in \mathbf{S}^n_{++}$ and is given by (\ref{eq:P}). This choice for $\mathbb{P}$ is motivated by the fact that it not only needs to guarantee that the system is stable but also needs to ensure that the adaptive laws are only a function of the state of agent $i$ and the state of its neighbors $j \in \mathcal{N}_i$.

With this choice of $\mathbb{P}$ and adjusting the control gains as in (\ref{eq:adaptation}), it can be shown using the properties of the Kronecker product \cite{Bellman1997IntroductionEdition}, (\ref{eq:P}b), and standard adaptive control arguments \cite{Narendra1989StableSystems}, that
\begin{equation}
\begin{split}
    \dot{V} = & \ \mathbf{e}^T \left[\left(I \otimes A_m^T\right) \mathbb{P} + \mathbb{P}\left(I \otimes A_m\right) -\right. \\
    & \ \left.\left(L \otimes \Theta^T B^T P\right) - \left(L \otimes P B \Theta\right)\right]\mathbf{e} \\
    = & \ -\mathbf{e}^T \left[\left(I \otimes Q\right) + \left(L \otimes \Theta^T B^T P\right) + \right. \\
    & \ \left. \left(L \otimes P B \Theta\right)\right]\mathbf{e} \\
    \leq & \ 0
\end{split}
\end{equation}

Since $V$ is positive definite and radially unbounded and $\dot{V}$ is negative semidefinite, then $\mathbf{e}$, $\tilde{\theta}^{(i)}_k \in \mathcal{L}_\infty$ for all $i \in \mathcal{V}$ and $k = \{1, 2, 3\}$. Furthermore, because $\dot{V} \leq 0$ we have that $\mathbf{e} \in \mathcal{L}_2$, and since $\mathbf{e}, \mathbf{\dot{e}} \in \mathcal{L}_\infty$, by Barbalat’s Lemma we are able to conclude that $\lim_{t\rightarrow\infty} \mathbf{e}(t) = 0$. 
\end{proof}

\subsection{Corollary 1:}\label{appendix B}

\begin{proof}
\
\begin{renum}
    \item Follows from the Lyapunov function (\ref{eq:Lyapunov}).
    \item Is a consequence of the asymptotic convergence of $\mathbf{e}(t)$ to zero. 
\end{renum}
\end{proof}

\subsection{Corollary 2:}\label{appendix C}

\begin{proof}
\
\begin{renum}
    \item Since by Theorem 2, $\tilde{\theta}^{(i)}_k \in \mathcal{L}_\infty$ for all $i \in \mathcal{V}$ and $k = \{1, 2, 3\}$, and because $||\omega(x_i(t),t)||_{\infty} \leq \bar{\omega}$ for all $t \geq t_0$, then by definition (\ref{eq:u_tilde_i}), $\tilde{u}_i \in \mathcal{L}_{\infty}$.
    \item Notice that (\ref{eq:error_dynamics}) is an LTI system with state $\mathbf{e}(t)$, bounded derivative $\mathbf{\dot{e}}(t)$, and input $\tilde{\mathbf{u}}(t)$, therefore it follows that $\lim_{t\rightarrow\infty} \tilde{\mathbf{u}}(t) = 0$ \cite{Annaswamy2021OnlineApproaches}. Because of the asymptotic convergence of $\tilde{\mathbf{u}}(t)$, for any $\varepsilon_u>0$, such $T_u$ exists. 
\end{renum}
\end{proof}


\subsection{Theorem 3}\label{appendix D}

\begin{proof}
Since h has relative degree one, the solution of the QP-GCBF filter, $\Delta_i^{\psi}$, is locally Lipschitz for $z_i \in \mathcal{S}_i$ for all $i \in \mathcal{V}$ (see \cite{Xu2015RobustnessControl}, Theorem 8). As a result, the conclusions of Corollaries 1 and 2 hold. The QP-GCBF filter also guarantees that the reference model solutions are safe. Since $z_i(t_0) \in \mathcal{S}_i$ for all $i \in \mathcal{V}$ and $\upsilon(t)>0$ for all $t \geq t_0$ in (\ref{eq:upsilon_t}), then
\begin{equation}\label{eq:safety_z_i}
    \frac{\partial h}{\partial z_i}^T\dot{z}_i \geq -\alpha_0 h(z_i) + \upsilon(t) > -\alpha_0h(z_i)
\end{equation}
which implies that $z_i(t) \in \mathcal{S}_i$ for all $t \geq t_0$. Let us know inspect the safety of the system by evaluating $\dot h(x_i)$:
\begin{equation}
    \begin{split}
        \frac{\partial h}{\partial x_i}^T\dot{x}_i & = \frac{\partial h}{\partial x_i}^T\left(Ax_i + B\left[\Lambda u_i + \Xi \omega\right]\right)\\
        & = \frac{\partial h}{\partial x_i}^T\left(Ax_i + B\left[\Lambda u_i + u_i^{\star} - u_i^{\star} + \Xi \omega\right]\right. + \\
        & \left.\quad A_m z_i - A_m z_i\right)\\
        & = \frac{\partial h}{\partial x_i}^T(Ae_i + A_m z_i + B\left[\Lambda \tilde{u}_i + \bar{u}_i\right])
    \end{split}
\end{equation}
After algebraic manipulation of (\ref{eq:safety_z_i}), we obtain that
\begin{equation}\label{eq:safety_x_i}
    \begin{split}
        \frac{\partial h}{\partial x_i}^T\dot{x}_i
        \geq & -\alpha_0 h(x_i) + \upsilon(t) - \alpha_0\left(h(z_i)-h(x_i)\right) + \\
        & \left(\frac{\partial h}{\partial z_i}^T-\frac{\partial h}{\partial x_i}^T\right)\left(A_m z_i + B\bar{u}_i\right) + \\
        & \frac{\partial h}{\partial x_i}^T(Ae_i + B\Lambda \tilde{u}_i)\\
        = & -\alpha_0 h(x_i) + \upsilon(t) + \eta(e_i(t),\tilde{u}_i,t)\\
        \geq & -\alpha_0 h(x_i) + \upsilon(t) - |\eta(e_i(t),\tilde{u}_i,t)|
    \end{split}
\end{equation}
where $\upsilon(t)$ is given by (\ref{eq:upsilon_t}). If \( x_i(t_0) \in \mathcal{S}_i \) for all \( i \in \mathcal{V} \), we will demonstrate that \( x_i(t) \in \mathcal{S}_i \) for all \( t \geq t_0 \) by analyzing two cases: (1) for \( t \in [t_0, t_0 + \hat{T}] \) and (2) for \( t > t_0 + \hat{T} \).
\begin{enumerate}
    \item For all $t \in [t_0, t_0 + \hat{T}]$, since $|e_i(t)|\leq E$ and $|\tilde{u}_i(t)| \leq \bar{U}_i(t)$ this implies that
    \begin{equation}
        |\eta(e_i(t),\tilde{u}_i,t)| \leq \bar{\eta}(E,\bar{U}_i(t),t)
    \end{equation}
    where $\bar{\eta}(E, \bar{U}_i(t),t)$ and $\bar{U}_i(t)$ are given by (\ref{eq:eta_bar}) and (\ref{eq:U_bar}), respectively.
    Therefore, by (\ref{eq:upsilon_t})
    \begin{equation}
        |\eta(e_i(t),\tilde{u}_i,t)| \leq \upsilon(t)
    \end{equation}
    and in turn
    \begin{equation}\label{eq:adaptation window safety}
        \frac{\partial h}{\partial x_i}^T\dot{x}_i
        \geq -\alpha_0 h(x_i), \quad \forall t \in [t_0, t_0 + \hat{T}]
    \end{equation}
    \item For all $t > t_0 + \hat{T}$, since $\hat{T} \geq \max\{T_e,T_u\}$, then $|e_i(t)|\leq ||\mathbf{e}(t)|| \leq \varepsilon_e$ and $|\tilde{u}_i(t)| \leq ||\tilde{\mathbf{u}}(t)|| \leq \varepsilon_u$, which implies that
    \begin{equation}
        |\eta(e_i(t),\tilde{u}_i,t)| \leq \bar{\eta}(\varepsilon_e,\varepsilon_u,t)
    \end{equation}
    Therefore, by (\ref{eq:upsilon_t})
    \begin{equation}
        |\eta(e_i(t),\tilde{u}_i,t)| \leq \upsilon(t)
    \end{equation}
    and in turn    \begin{equation}\label{eq:long term safety}
        \frac{\partial h}{\partial x_i}^T\dot{x}_i
        \geq -\alpha_0 h(x_i), \quad \forall t >t_0 + \hat{T}
    \end{equation}
\end{enumerate}
The inequalities in (\ref{eq:adaptation window safety}) and (\ref{eq:long term safety}) imply that 
    \begin{equation}\label{eq:overall safety}
            \frac{\partial h}{\partial x_i}^T\dot{x}_i
            \geq -\alpha_0 h(x_i), \quad \forall t \geq t_0
    \end{equation}
\begin{renum}
    \item Inequality in (\ref{eq:overall safety}) implies that $x_i(t) \in \mathcal{S}_i$ for all $i \in \mathcal{V}$ and all $t \geq t_0$, if $x_i(t_0) \in \mathcal{S}_i$ for all $i \in \mathcal{V}$.
    \item Since $S_i$ is a connected set, and by Corollary 1(ii) and Corollary 2(ii), for any $\varepsilon$, there exists a finite time $T$, such that $\upsilon(t) \leq \varepsilon$. Then, with an appropriate choice of $\varepsilon_e$, $\varepsilon_u$, and $\hat{T} \geq T$, the set $\tilde{\mathcal{S}}_i(\upsilon(t)) \subset \tilde{\mathcal{S}}_i(\varepsilon)$ remains connected for all $t > t_0 + \hat{T}$. Therefore, if $x_i^{\star} \in \tilde{\mathcal{S}}_i(\varepsilon)$, then $x_i(t) \rightarrow x_i^{\star}$ as $t \rightarrow \infty$ for all $i \in \mathcal{V}$. 
\end{renum}




\end{proof}

\end{document}